\documentclass[11pt]{article}
\pdfoutput=1

\usepackage{euscript}
\usepackage{amssymb}
\usepackage{amsfonts}
\usepackage{amsbsy}
\usepackage{amsmath}
\usepackage{epsfig}
\usepackage{amsthm}
\usepackage{amscd}
\usepackage{amstext}
\usepackage{verbatim}

\def\ben{\begin{equation}}
\def\een{\end{equation}}

\let\a=\alpha  \let\g=\gamma  \let\e=\varepsilon
   \let\k=\kappa

\let\w=\omega  \let\D=\Delta

\let\pa=\partial
\def\be{\begin{equation}}
\def\ee{\end{equation}}
\def\beq{\begin{equation}}
\def\eeq{\end{equation}}
\def\ba{\begin{array}}
\def\ea{\end{array}}

\def\dalemb#1#2{{\vbox{\hrule height .#2pt
       \hbox{\vrule width.#2pt height#1pt \kern#1pt
               \vrule width.#2pt}
       \hrule height.#2pt}}}

\newcommand{\bea}{\begin{eqnarray}}
\newcommand{\eea}{\end{eqnarray}}

\newcommand{\Tr}{{\rm Tr} }

\def\R{{{\Bbb R}}}

\def\ocal{{\mathcal{O}}}

\begin{document}

\begin{center}

{ \LARGE {\bf Entropy balance in holographic superconductors}}

\vspace{1cm}

{\large Sean A. Hartnoll$^\flat$ and Razieh Pourhasan$^\sharp$

\vspace{0.7cm}

{\it $^\flat$ Department of Physics, Stanford University, \\
Stanford, CA 94305-4060, USA \\}

\vspace{0.3cm}

{\it $^\sharp$ Department of Physics \& Astronomy, University of Waterloo, \\
Waterloo, Ontario N2L 3G1, Canada} }

\vspace{1.6cm}

\end{center}

\begin{abstract}

In systems undergoing second order phase transitions, the temperature integral of the specific heat over temperature from zero to the critical temperature is the same in both the normal and ordered phases. This entropy balance relates the critical temperature to the distribution of degrees of freedom in the normal and ordered states. Quantum criticality and fractionalization can imply an increased number of low energy degrees of freedom in both the normal and ordered states. We explore the r\^ole of entropy balance in holographic models of superconductivity, focussing on the interplay between quantum criticality and superconductivity. We consider models with and without a ground state entropy density in the normal phase; the latter models are a new class of holographic superconductors. We explain how a normal phase entropy density manifests itself in the stable superconducting phase.

\end{abstract}

\pagebreak
\setcounter{page}{1}

\section{Introduction}

Perhaps the central question concerning superconductivity is the most obvious one: when can the critical superconducting temperature become high? What circumstances favor quantum mechanical pairing instabilities against the stabilizing effects of thermal fluctuations? The theorist might like to identify aspects of these big questions that can be abstracted as much as possible from the specifics of concrete materials.

As we will shortly recall, systems undergoing second order phase transitions obey a type of entropy balance: the integral of the specific heat over temperature, from zero temperature to the critical temperature, has to be the same when evaluated in the (stable) superconducting state and in the (unstable) normal state. This gives a formula that relates the critical temperature to the difference in the distribution of degrees of freedom between the normal and superconducting states.
Read in the most na\"ive way, it suggests that having high critical temperatures is correlated with the normal state having many more low energy degrees of freedom than the superconducting state. For conventional superconductors the entropy balance is well understood, as we will briefly review: the superconducting state is gapped, the number of `surplus' low energy degrees of freedom in the normal state are proportional to this gap, and consequently $T_c$ is also proportional to the gap.

Unconventional superconductors have unconventional normal states. There is broad evidence for the notion that the normal states are unconventional because the low energy fermionic degrees of freedom are strongly interacting with collective critical bosonic excitations. See e.g. \cite{subir} for a suggestive overview of several different compounds. 
In terms of the entropy balance, additional low energy degrees of freedom in the normal state bodes well for obtaining a higher critical temperature. It is important to have concrete, controlled models where the entropy balance can be explicitly worked out in such cases. For instance, as we shall see, it can occur that the critical degrees of freedom are not all gapped out in the superconducting state. In this paper we shall see how the entropy balance pans out in holographic models of superconductivity. As we shall also briefly review, these are theoretical models for the emergence of superconductivity from fractionalized phases of matter, wherein the charge carriers are strongly coupled to deconfined gauge fields.

An important impetus for our study were the recent careful measurements of the temperature dependence of the specific heat in Sr$_3$Ru$_2$O$_7$ \cite{andy}. The evidence suggests that, in a suitably tuned magnetic field, this material is a two (space) dimensional quantum critical metal that undergoes nematic ordering at low temperatures, as reviewed in \cite{andy2}. The general considerations we have described above apply equally to nematic as well as superconducting instabilities. Among the several dramatic discoveries in \cite{andy} is that, as the temperature is lowered through the critical nematic temperature, the specific heat over temperature continues to increase in the nematic phase as the system is cooled. It seems natural to think that this increase is made possible by the persistence of quantum critical degrees of freedom in the ordered state. In the models studied in this paper, which are presumably not related to Sr$_3$Ru$_2$O$_7$, we will precisely explore the effects of quantum criticality on the redistribution of degrees of freedom as the system undergoes an ordering transition.

A second motivation for this work is that, as we will recall below, many of the simplest models of holographic superconductivity generically have the property that the normal state of the system is locally quantum critical and exhibits a zero temperature ground state entropy density. This state of affairs is usually considered to be `cured' by the onset of superconductivity in the sense that the superconducting state at zero temperature has vanishing entropy density. Nonetheless, we will see that entropy balance implies that the normal and superconducting states are not independent; the zero temperature entropy density of the normal state can have important consequences for the thermodynamics of the superconducting state. For comparison, we will study holographic models of superconductivity with and without a ground state entropy in the normal state.

Finally, we have also been inspired by recent applications of the notion of ``quantum order-by-disorder'' to finite density systems, namely, Fermi liquids in the vicinity of ferromagnetic instabilities \cite{aa,bb,cc}.  The idea is that the low energy degrees of freedom associated with disordering the original ferromagnetic phase can, as the quantum critical point is approached, favor new orderings. Technically this is described in \cite{aa,bb,cc} using second order perturbation theory about a mean field treatment. One would like to have a general model-independent formulation, and furthermore not tied to a perturbative understanding of the difference between classical and quantum effects, of what quantum order-by-disorder means. The notion of entropy balance that we will shortly review in detail has the flavor of a generalized notion of quantum order-by-disorder, as it can capture general features relating to the interplay of quantum criticality and ordering.

\section{Second order phase transitions and entropy balance}

At a second order phase transition, the free energy and entropy densities of the superconducting and normal states are both equal at the critical temperature
\be\label{eq:fs}
f_N(T_c) = f_S(T_c) \,, \qquad s_N(T_c) = s_S(T_c) \,.
\ee
The specific heat, in contrast, will jump discontinuously at the critical temperature and, just below $T_c$, will be larger in the superconducting state
\be\label{eq:c}
c_N(T_c) < c_S(T_c) \,.
\ee
This inequality expresses the fact that the superconducting state is preferred for $T < T_c$. The above statements are strictly true if the transition has a mean field form. This will be the case for all of the models considered in this paper.

From the equality of the entropies at the critical temperature, and from the definition of the specific heat as $c/T = \pa s/\pa T$, one obtains the following `sum rule' relating the integrals of the specific heat in the superconducting and normal phases
\be\label{eq:sum}
\int_0^{T_c} \frac{c_N(T)}{T} dT + s_N(0) = \int_0^{T_c} \frac{c_S(T)}{T} dT \,.
\ee
On the left hand side we have allowed for a finite entropy density at zero temperature in the normal state, as this will be present for some of the models we consider. The relation (\ref{eq:sum}) imposes an `entropy balance' constraint that keeps track of the degrees of freedom, even though the
excitations of the system can be drastically re-organized due to the onset of superconductivity.

The entropy balance equation (\ref{eq:sum}) suggests the following observation. Suppose we wish to tweak the system to increase the critical temperature. How might we do it? The inequality (\ref{eq:c}) tells us that the contribution from near the upper limit of the integrals,
just below $T_c$, is larger in the superconducting state. Therefore, if we increase $T_c$, the right hand side of equation (\ref{eq:sum}) becomes larger. To maintain the entropy balance it follows that at temperatures well below the critical temperature we might either (a) increase the low temperature specific heat of the normal state or (b) decrease the low temperature specific heat of the superconducting state. We could of course do both. A, somewhat na\"ive, physical conclusion would seem to be that to increase the critical temperature we should try to increase the low energy degrees of freedom of the normal state relative to the low energy degrees of freedom of the superconducting state. An alternative way to increase the critical temperature is to decrease the jump in the heat capacity (\ref{eq:c}) at $T_c$. This paper will be an investigation of such correlations in holographic models of superconductivity.

\section{BCS theory}

\begin{figure}[h]
\begin{center}
\includegraphics[height=180pt]{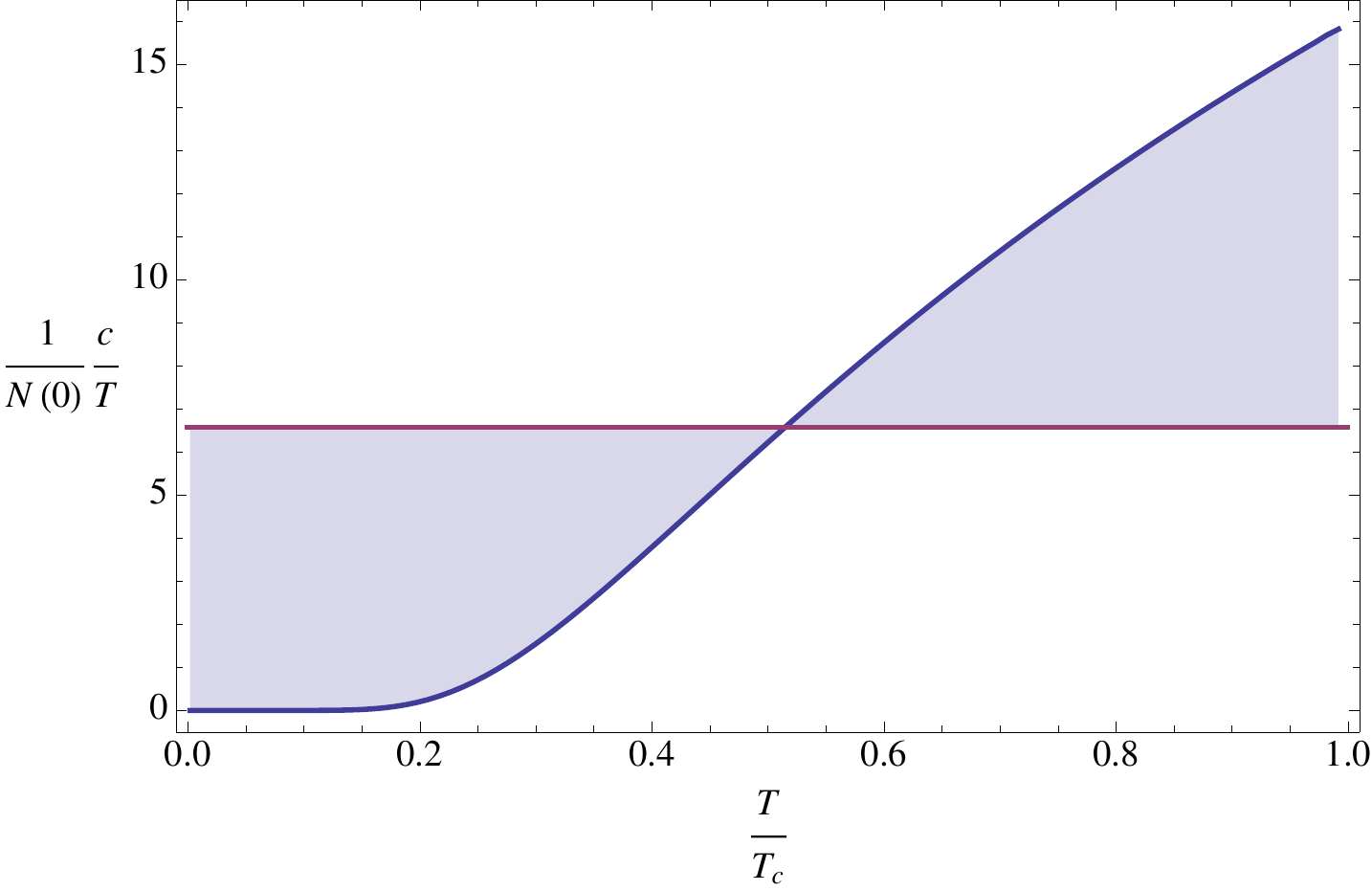}\caption{Entropy balance in BCS theory. Specific heat over temperature for the normal and superconducting states below $T_c$. The area between the two curves integrates to zero. \label{fig:bcsovert}}
\end{center}
\end{figure}

To contextualize, we will very briefly review how the entropy balance works out in the BCS theory of superconductivity.
We will only consider a mean field treatment here.
The key formula is of course the famous gap equation
\be
\frac{1}{g N(0)} = \int_0^{\w_D} \frac{d\e}{E} \tanh \frac{E}{2 T} \,, \qquad E^2 = \e^2 + \Delta_T^2 \,.
\ee
Here $g$ is the electron-phonon coupling and $N(0)$ the density of states at the Fermi surface. This equation immediately gives the zero temperature gap and the critical temperature in terms of the gap
\be\label{eq:gap}
\Delta_0 = 2 \w_D e^{-1/g N(0)} \,, \qquad T_c \approx 0.57 \Delta_0 \,.
\ee
The specific heat of the superconducting and normal phases are (e.g. \cite{BCS})
\be
c_S = N(0) \int_0^{\w_D} d\e \, \text{sech}^2 \frac{E}{2T} \cdot \left(\frac{E^2}{T^2} - \frac{E}{T} \frac{dE}{dT} \right) \,, \qquad c_N = \frac{2 \pi^2}{3} N(0) T \,.
\ee
From this formula we obtain the low temperature specific heat in the superconducting phase
\be\label{eq:lowT}
\frac{c_S}{T} = \sqrt{8 \pi} N(0) \, (\D_0/T)^{5/2} \, e^{-\D_0/T} + \cdots \,.
\ee
Inspecting equations (\ref{eq:gap}) and (\ref{eq:lowT}) we see that the na\"ive intuition from above is realized: Increasing $\Delta_0$ increases $T_c$ and simultaneously decreases the degrees of freedom in the superconducting state at low temperatures. Plotting the specific heat of the two phases in figure \ref{fig:bcsovert}, for small $g N(0)$, we can see exactly how the entropy balance works out. \\

The main feature in the BCS specific heat plot is the gap in the superconducting phase at low temperatures. In the holographic models that we will be studying in the following, the normal state is quantum critical and not all of the quantum critical degrees of freedom will be gapped in the superconducting state. We wish to use holographic models to examine the interplay between criticality in the normal and ordered states and the critical temperature.

\section{Holographic superconductors}

\subsection{The condensed matter physics of holographic superconductors}

Holographic superconductors are a gravitationally dual description of the onset of superconductivity from certain fractionalized phases of matter. By fractionalized phase we mean that, somewhat heuristically, the electric charge may be thought of as being carried by degrees of freedom that are charged under an emergent large $N$ gauge symmetry. These could be, for instance, $N\times N$ matrix-valued fermion fields $\Psi$ or boson fields $\Phi$. This gauge symmetry is in addition to the global $U(1)$ `electric' charge. The superconducting instability occurs when it becomes favorable for the charge to be carried by short gauge-invariant bosonic operators, such as $\Tr \left( \Psi \Psi \right)$ or $\Tr \left( \Phi \Phi \right)$. We will need two different flavors of fermionic field in order to make the trace of the product nonvanishing.  These are not Cooper pairs of the UV gauge invariant `physical' electrons, but rather pairs of low energy fractionalized excitations. The physical electrons themselves are given by operators of the schematic form $\Tr \left(\Psi \Phi \right)$. The corresponding more conventional Cooper pairs would then correspond to the `double trace' operator $\Tr \left(\Psi \Phi \right) \Tr \left(\Psi \Phi \right)$. Such pairing can also be described holographically \cite{Hartman:2010fk}.

The fractionalized phase is gravitationally described by a charged black hole horizon. The formation of the `meson' Cooper pairs described above is captured by the expulsion of charge from the horizon into a charged Bose-Einstein condensate outside the black hole. In the zero temperature limit, often all of the charge is expelled, yet a (neutral) horizon remains. This indicates that electrically neutral critical excitations, such as deconfined gauge fields, have survived the onset of superconductivity. This phenomenon is reminiscent of the Landau damping of transverse gauge fields by a charge density, in which the transverse modes are not screened. The presence of critical modes in both the normal and superconducting states at low temperatures will be a recurring theme in this paper.

A more detailed discussion of the above ideas appears in \cite{Hartnoll:2011fn}, along with references to the original literature.

The discussion above shows that two processes occur simultaneously at $T_c$ in holographic superconductors. The global $U(1)$ symmetry is spontaneously broken by a condensate and furthermore the fractionalized charge starts to `confine' into meson-like degrees of freedom. This is a little different to the emergence of superconductivity from a fractionalized phase in e.g. \cite{senthil}, where the analogue of the gauge-charged boson $\Phi$ condenses. In that case the global $U(1)$ is spontaneously broken and the emergent gauge symmetry is Higgsed. Current holographic superconductor models do not appear to involve Higgsing of the $SU(N)$ gauge symmetry. Pairing analogous to $\Tr \left( \Psi \Psi \right)$ was not accessible in \cite{senthil} because the fractionalized fermionic excitations were electrically neutral. Other models of phases with emergent gauge fields such as the `doublon metal' do have pairing instabilities in which the condensate is gauge-neutral but electromagnetically charged \cite{doublon}. Various theories with similar phases, involving the simultaneous onset of confinment in the charged sector and superconductivity, are described in \cite{Huijse:2011hp}. These are very much analogous to holographic superconductivity.

\subsection{Models with a normal state entropy density at $T=0$}

We will proceed to study two classes of holographic superconductors. The first, simplest class, will have the curious feature that the normal state is locally quantum critical and has a zero temperature entropy density. Perhaps unsurprisingly, we will see this fact, the zero temperature entropy density, strongly colors the way in which the entropy balance is realized in these theories. In the following subsection we will move on to models that display a more generic quantum criticality in their normal states.

A simple class of models of holographic superconductors have the bulk Lagrangian
\be\label{eq:EMH}
{\mathcal L} = \frac{1}{2 \k^2} \left(R + \frac{6}{L^2} \right) - \frac{1}{4 e^2} F_{\mu\nu} F^{\mu\nu} - \frac{1}{\k^2} \left( |\nabla \phi - i A \phi|^2 + m^2 |\phi|^2 + V(|\phi|) \right) \,.
\ee
We will work with a 3+1 dimensional bulk, and hence a 2+1 dimensional dual quantum field theory.
We will very briefly review the properties of the normal state of such models, with $\phi = 0$. A more pedagogical exposition with the same conventions as we are using here can be found in \cite{Hartnoll:2011fn}.
The normal state is holographically described by the planar Reissner-Nordstr\"om-AdS black hole, with metric and Maxwell potential
\be
ds^2 = \frac{L^2}{r^2} \left(- f(r) dt^2 + \frac{dr^2}{f(r)} + dx^2 + dy^2 \right) \,, \qquad A = \mu \left(1 - \frac{r}{r_+} \right) dt \,.
\ee
Here $\mu$ will be the chemical potential of the dual field theory. The metric function is
\be
f(r) = 1 - \left(1 + \frac{r_+^2 \mu^2}{2 \g^2} \right) \left(\frac{r}{r_+}\right)^{3} + \frac{r_+^2 \mu^2}{2 \g^2} \left(\frac{r}{r_+}\right)^{4} \,.
\ee
We introduced the dimensionless ratio of the Newtonian and Maxwell couplings
\be
\g^2 = \frac{e^2 L^2}{\k^2} \,.
\ee
The horizon radius $r_+$ determines the temperature as
\be
T = \frac{1}{4 \pi r_+} \left(3 - \frac{r_+^2 \mu^2}{2 \g^2} \right) \,.
\ee
The entropy density and specific heat of the solution are
\be
s_N = \frac{L^2}{\k^2} \frac{2 \pi}{r_+^2} \,, \qquad c_N = T \frac{\pa s}{\pa T} = \frac{L^2}{\k^2} \frac{1}{r_+^2} \frac{8 \pi^2 T r_+}{3 - 2 \pi T r_+ } \,.
\ee
The temperature derivative has been taken with $\mu$ fixed.

This class of models has the property that at zero temperature, the near horizon geometry of the extremal Reissner-Nordstr\"om-AdS black hole is $AdS_2 \times \R^2$. This implies that the low energy physics develops an emergent local quantum criticality, with scaling symmetry $\{t \to \lambda t, \, x \to x\}$ in which space does not scale. Associated with this symmetry is the widely discussed zero temperature entropy density
\be
s_N(0) = \frac{\pi \mu^2}{3 e^2} \,.
\ee
Such an entropy density is not a necessary consequence of local criticality \cite{Hartnoll:2012wm}, but does arise robustly in holographic models described at zero temperature by a regular extremal event horizon.

In these models, the Reissner-Nordstr\"om solution can become unstable towards condensation of $\phi$ below a critical temperature $T_c$ \cite{Gubser:2008px, Hartnoll:2008vx, Hartnoll:2008kx}. The critical temperature itself depends on the mass $m^2$ of the scalar field as well as ratio $\g$ \cite{Denef:2009tp}. Despite the strong interactions of the dual quantum field theory, in which operators certainly have large anomalous dimensions, the finite temperature phase transition is mean field in nature because the large $N$ limit suppresses fluctuations \cite{Anninos:2010sq}. To discuss the superconducting state, we need to make a choice for the scalar potential and mass.

A convenient set of models to consider have \cite{Gubser:2009cg}
\be\label{eq:mV}
m^2 L^2 = - 2 \,, \qquad V(|\phi|) = \frac{u}{2 L^2} |\phi|^4 \,.
\ee
This choice of the mass corresponds to taking the operator $\ocal$ dual to $\phi$ to have scaling dimension $\Delta = 2$ in the conformally invariant UV fixed point theory. The choice of potential ensures that at zero temperature in the superconducting phase the scalar field does not run off to infinity in the IR geometry.\footnote{While the slightly simpler case of positive mass squared and no potential also admits superconducting Lifshitz solutions in the IR at zero temperature \cite{Gubser:2009cg, Horowitz:2009ij}, with a stabilized scalar, fluctuations about these backgrounds always include modes with complex scaling dimensions and hence the geometries are unstable \cite{Hartnoll:2011pp, Edalati:2011yv}. Presumably, the true IR geometry in these cases has a runaway scalar.} This will make it easier to determine the low temperature specific heat of the system.
In fact, the near horizon geometry at zero temperature has a scaling invariance characterized by a dynamical critical exponent $z$
\be\label{eq:lif}
\frac{1}{L^2}  ds^2 = - \frac{dt^2}{r^{2z}}  + g_L \frac{dr^2}{r^2} + \frac{dx^2 + dy^2}{r^2} \,, \quad A = \g \, h_L \frac{dt}{r^z} \,, \quad \phi = \phi_L \,.
\ee
Here $\{g_L, h_L, \phi_L\}$ are constants. The exponent and stabilized value of the scalar field satisfy
\bea
\g^2 \phi_L^2 (4 + z + z^2) + z (u \phi_L^4 - 4 \phi_L^2 - 6) & = & 0 \,, \label{eq:aa} \\
6 \g^2 \phi_L^2+ z \left(u \phi_L^4 (3+z) - 2 \phi_L^2 (4+z) - 6
\right) & = & 0 \,. \label{eq:bb}
\eea
Furthermore $h_L^2 = (z-1)/z \,, g_L = z/(\g^2 \phi_L^2)$. Note that $z>1$.

%
%
%, however, For fixed $z$ and
%$\gamma$ we get
%\bea\label{eq:uphi}
%&&u=\frac{z^3 \gamma ^2
%\left(2+\gamma ^2\right)+z^2 \left(-4+2 \gamma ^2+\gamma
%^4\right)+z \gamma ^2
%\left(8+\gamma ^2\right)-3 \gamma ^4}{6 z^2}\\
%&&\phi_L=\sqrt{\frac{6z}{z^2 \gamma ^2+2 z \left(-1+\gamma
%^2\right)+3 \gamma ^2}}
%\eea
%which gives some constraint on
%$\gamma$ in order to avoid complex values for $\phi_L$.

The `Lifshitz' \cite{Kachru:2008yh} scaling of the near horizon
metric (\ref{eq:lif}) at zero temperature implies that at low temperatures
\be\label{eq:clowT}
c_S(T) \sim T^{2/z} \,.
\ee
This follows from dimensional analysis. Thus the larger $z$ is, the more degrees of freedom are present in the superconducting phase at low temperatures. Because $z$ is a directly physical quantity, we will parametrize the theories by $\{z,\g\}$ rather than $\{u,\g\}$. Given $z$ and $\g$ it is easy to obtain the corresponding $u$ from 
(\ref{eq:aa}) and (\ref{eq:bb}). The low temperature specific heat (\ref{eq:clowT}) is qualitatively different from the gapped BCS formula (\ref{eq:lowT}). As discussed above, it indicates that the critical modes in the normal state are not all gapped by the superconducting condensate in these models.

The full spacetime, at zero and finite temperature, takes the
general form
\be\label{eq:spacetime} \frac{1}{L^2} ds^2 = -
f(r)dt^2 + g(r) dr^2 + \frac{dx^2 + dy^2}{r^2} \,, \quad A = \g \,
h(r) dt \,, \quad \phi = \phi(r) \,.
\ee
In these coordinates $r
\to 0$ will be the $AdS_4$ conformal boundary. This ansatz leads
to the following equations of motion
\bea\label{eq:EOM}
 r \phi'^2 + \frac{f'}{2f} + \frac{g'}{2g} + \frac{2}{r} + \g^2 r \, \frac{g h^2 \phi^2}{f} & = & 0 \,, \\
 r \frac{h'^2}{f} - \frac{f'}{f} + \frac{g'}{g} + \frac{6}{r} - r g (6 + 4 \phi^2 - u \phi^4) & = & 0 \,, \\
 h''  - h' \left(\frac{2}{r} + \frac{f'}{2f} + \frac{g'}{2g} \right) - 2 \g^2 g h \phi^2 & = & 0 \,, \\
 \phi'' - \left(\frac{2}{r} - \frac{f'}{2f} + \frac{g'}{2g} \right) \phi' + \left(2g + \g^2 \frac{g h^2}{f} \right) \phi - u g \phi^3 & = & 0 \,.
\eea
The strategy to obtain the specific heat as a function of temperature is now as follows. We must first find a solution to the above equations such that near the boundary $\phi \sim \langle \ocal \rangle r^2$ is purely normalizable. Away from the boundary, the solution should end at a regular black hole horizon at $r=r_+$. Near the boundary we can read off the speed of light from $f \sim c_o^2 r^{-2}$. This is important because we must rescale the time coordinate so that all theories are compared with the same speed of light. By varying data at the horizon, we will find a one parameter family of solutions, from which we can obtain the specific heat using the following expressions for the temperature and entropy density
\be
T = \frac{1}{4 \pi c_o} \left| \frac{df}{dr} \frac{d(g^{-1})}{dr} \right|^{1/2}_{r=r_+} \,, \qquad s = \frac{L^2}{\k^2} \frac{2 \pi}{r_+^2} \,.
\ee
The family of solutions will correspond to temperatures ranging from $T=0$ to $T=T_c$.
It is important to compute dimensionless quantities; in fact we should obtain $c/(\mu T)$ as a function of $T/\mu$, where the chemical potential is read off from the near boundary behaviour of the Maxwell potential: $h \sim c_o \mu/\g$. We are imagining varying the temperature with the chemical potential held fixed.

As usual, the equations are solved numerically by integrating out from the horizon and shooting to impose the boundary condition $\phi \sim \langle \ocal \rangle r^2$. In general multiple families of solutions will be found and it is important to take the stable one, with the highest $T_c$. As noted in \cite{Gubser:2009cg}, the existence of a Lifshitz scaling solution (\ref{eq:lif}) does not guarantee that this solution arises as the IR limit of an asymptotically AdS spacetime. The condition for this to occur is found by perturbing the solutions away from Lifshitz
\be\label{eq:pertL}
f=\frac{1}{r^{2z}}\left(1+f_1r^{\alpha} \right)\,,\quad
g=\frac{g_L}{r^{2}}\left(1+g_1r^{\alpha} \right)\,,\quad
h=\frac{h_L}{r^{z}}\left(1+h_1r^{\alpha} \right)\,,\quad
\phi=\phi_L\left(1+\phi_1r^{\alpha} \right) \,,
\ee
and requiring that there are at least two values of $\a < 0$, such that the perturbations are irrelevant, i.e. grow towards the UV, and can be tuned. Substituting the above expansion into the equations (\ref{eq:EOM}) we will get five allowed nonzero exponents $\a$; three are always positive and the other two can be negative depending on the value of $\gamma$. See \cite{Gubser:2009cg} for a plot. In this way, a portion of possible values of $\{z,\g\}$ do not in fact arise. We also require that $\a$ be real, as complex $\a$ indicates an instability \cite{Hartnoll:2011pp, Edalati:2011yv}.

We saw in (\ref{eq:clowT}) that the number of degrees of freedom at the lowest temperatures in the ordered phase is determined by $z$. However, the critical temperature $T_c$ in fact only depends on $\g$ and not on $z$ at all. We see this by recalling that $T_c$ is determined by a linearized analysis, where the charged scalar field is small. In this limit the nonlinear terms in the potential $V(|\phi|)$ in (\ref{eq:EMH}) drop out, and so, at fixed mass $m$, the equations of motion for the scalar field only depend on $\g$. The critical temperatures as a function of $\g$ for a range of masses were found in \cite{Denef:2009tp}, using slightly different conventions to the present paper. In particular it was shown that $T_c/\mu$ increased with increasing $\g$, as this may be thought of as the effective charge of the scalar field.

\begin{figure}[h]
\begin{center}
\includegraphics[height=200pt]{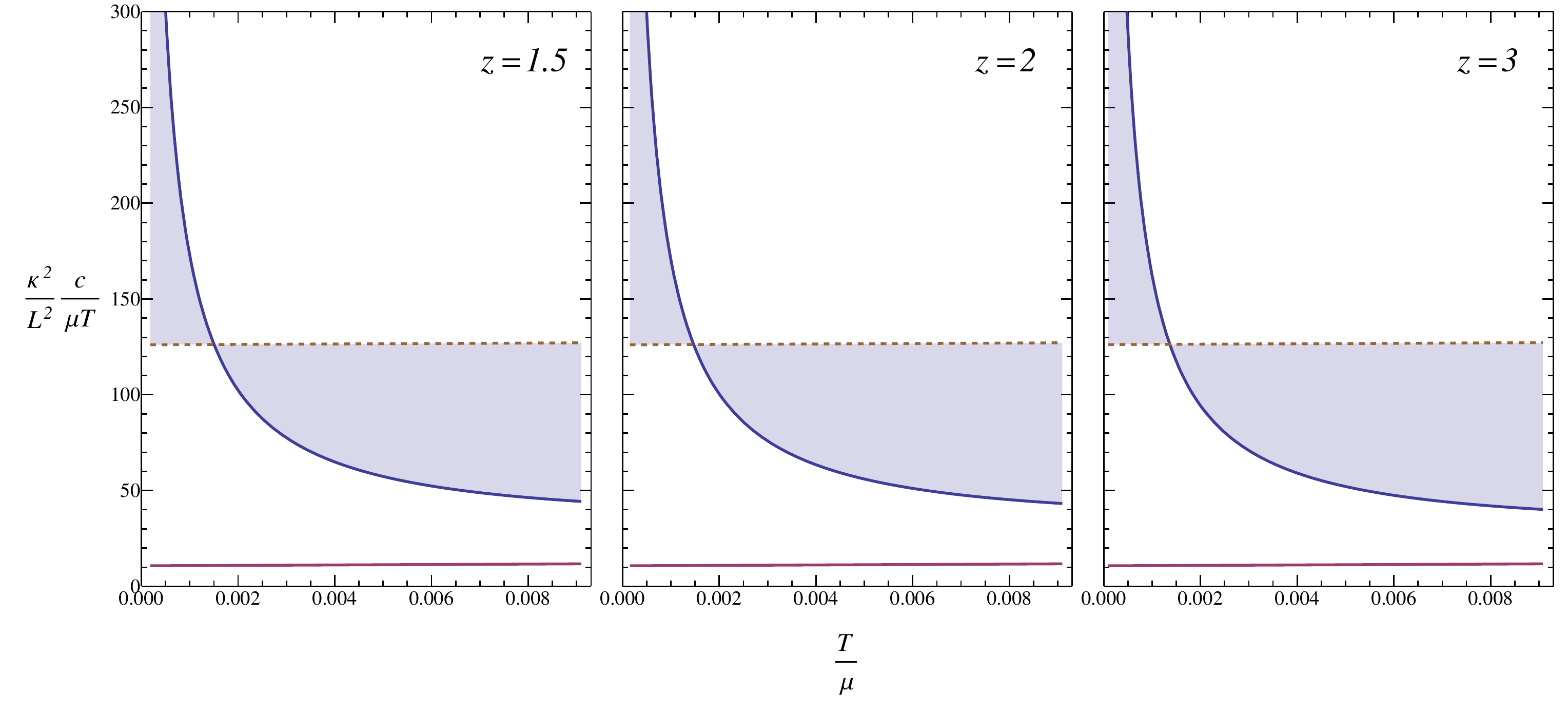}\caption{Specific heat over temperature as a function of temperature
for the normal (red) and superconducting (blue) phases below $T_c$ and with $\gamma=1$. The plots are for three different values of $z$, for a model with a nonzero ground state entropy density $s_N(0)$. The shaded region between the blue curve and the dashed yellow line, indicating $c_N/T + s_N(0)/T_c$, integrates to zero area.
\label{fig:covertg1}}
\end{center}
\end{figure}

The independence of $T_c$ and $z$ therefore seems to contradict our intuition that, if the normal state is kept fixed, there should be an anti-correlation between the critical temperature and the number of low energy degrees of freedom in the superconducting state. However, this is not the whole story. In figure \ref{fig:covertg1} below we plot 
$c/T$ against $T < T_c$ for the relatively small value of $\gamma = 1$, and for three values of $z = 1.5,2,3$.\footnote{The Lifshitz solution with $\gamma=1$ and $z=3$ is not in fact the true IR in this case, as there are operators with complex scaling dimension, i.e. $\a$ in (\ref{eq:pertL}), indicating an instability. This is not crucial for the points we are making here, and will not be the case in any other solutions we consider.}
The first important feature of these three plots is that they are indistinguishable. We knew that $T_c$ would be the same, as it only depends on $\g$, but the low temperature behavior should have depended strongly on $z$; from (\ref{eq:clowT}) we see that $c_S/T$ goes to zero for $z=1.5$, to a constant for $z=2$, and diverges for $z=3$.

In this paragraph and the following, we explain how the plots in figure \ref{fig:covertg1} can be nicely understood from entropy balance, despite being dramatically different from the BCS curve of figure \ref{fig:bcsovert}.
We see that for this value of $\g$, $T_c/\mu$ is small, of order $10^{-2}$. Correspondingly, we also see that the superconducting state has a large number of degrees of freedom at low temperatures. So many, in fact, that it has completely swamped out the $z$ scaling behavior. If we proceed to \emph{extremely} low temperatures, five or six orders of magnitude below $T_c$, we will eventually see the expected scaling. The low $T_c$ combined with the independence of $z$ from $T_c$ has forced the effects of $z$ to be pushed down to (presumably physically irrelevant) very low temperatures.

There is a further reason why the superconducting state has so many degrees of freedom at low temperatures in these plots. This is the ground state entropy density of the normal state. According to the entropy balance, these degrees of freedom must reappear somewhere in the superconducting state. From the entropy balance formula (\ref{eq:sum}) we see that the integral of $c_S(T)/T$ must equal the integral of $c_N(T)/T + s_N(0)/T_c$. A nonzero $s_N(0)$ and a low $T_c$ therefore push up the curve against which the superconducting phase degrees of freedom must balance. This curve is shown as a dashed yellow line in figure \ref{fig:covertg1}. As we recalled above, immediately below $T_c$ the superconducting phase must have a higher specific heat than the normal phase. However, in the presence of a ground state entropy density, this is not sufficient for the specific heat to need to decrease at low temperatures. In figure \ref{fig:covertg1} we see that the specific heat of the superconducting phase is below the dashed yellow line at $T_c$ and is therefore forced to increase as the temperature is lowered, in order to make the entropy balance work. The ground state entropy of the normal state is therefore directly responsible for the temperature dependence of the specific heat in the superconducting state being markedly different to the prototypical BCS curve  of figure \ref{fig:bcsovert}. An increase in $c/T$ as the temperature was lowered below $T_c$ in the ordered phase was one of the anomalous experimental features found in \cite{andy}, although no evidence was found there for a ground state entropy density. We will return to experiments in the discussion.

Note that $c_N/T$ is a constant in figure \ref{fig:covertg1} not because the normal state is a Fermi liquid, it is not, but because the critical temperature is so low in this case that the specific heat of the normal state does not vary over the range $0 < T < T_c$.

\begin{figure}
\begin{center}
\includegraphics[height=570pt]{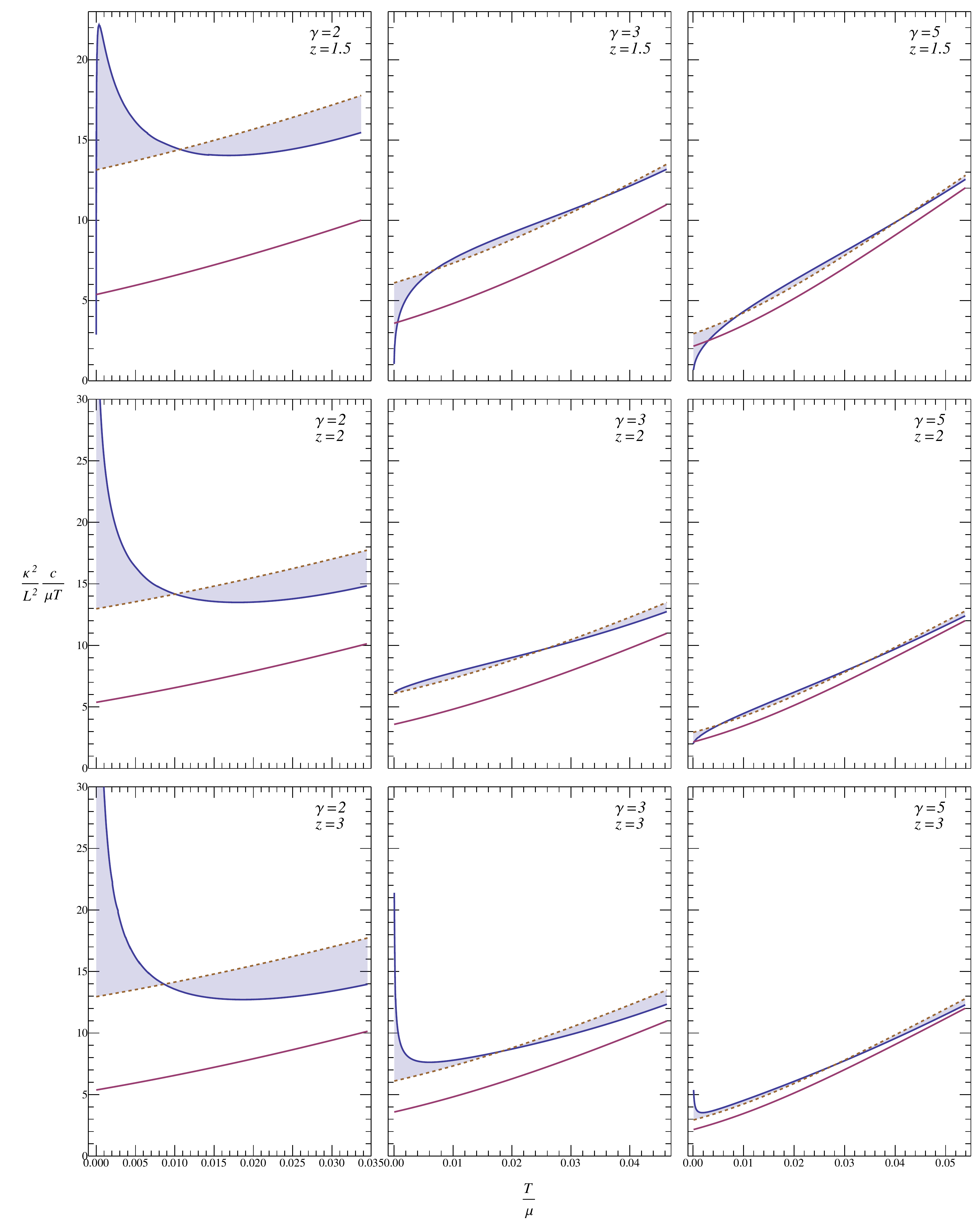}\caption{Specific heat over temperature as a function of temperature
for the normal (red) and superconducting (blue) phases below $T_c$. This is for a model with a ground state entropy density $s_N(0)$. Three values of $\g$ and $z$ are shown.
The shaded region between the blue curve and the dashed yellow line, indicating $c_N/T + s_N(0)/T_c$, integrates to zero area.
 \label{fig:covert}}
\end{center}
\end{figure}

Let us now increase the critical temperature by increasing $\g$. Figure \ref{fig:covert} shows $c/T$ against $T/\mu$. We again take $z=1.5,2,3$ but now with $\g = 2,3,5$.
The main feature of these plots is that as $T_c$ is increased, the anticipated low temperature scaling governed by $z$ becomes visible. We have checked that the numerics reproduce (\ref{eq:clowT}). This is very much consistent with our discussion in the previous paragraphs. As $T_c$ is increased the system (i) has a wider range of temperature over which to balance the entropy and therefore the $z$ scaling does not get swamped out and (ii) the effect of the normal state zero temperature entropy density on the entropy balance is weakened according to $s_N(0)/T_c$.

For $z \leq 2$, the specific heat over temperature must still raise in the superconducting state as the temperature is lowered below $T_c$, in order to cross the dashed line and achieve entropy balance, but then decreases at low temperatures as dictated by $z$. For $z > 2$, $c_S/T$ is monotonic.

In this subsection we have seen how important features of the thermodynamics of the superconducting phase in these models can be understood by reference to the thermodynamics of the normal state. We saw that the nonzero $s_N(0)$ has a substantial impact on the specific heat in the superconducting phase, in some regimes swamping out the quantum critical physics. In the following subsection we turn to a quantum critical holographic model without a zero temperature entropy density in the normal state.

\subsection{Models with critical normal states with $z_N$ finite}

There are several ways in which to obtain a normal state with finite $z_N$ criticality. We use $z_N$ to denote the dynamical critical exponent of the normal state, in order to differentiate it from the $z$ of the superconducting state that we discussed at some length in the previous subsection. The simplest setup seems to be, following \cite{Taylor:2008tg, Goldstein:2009cv}, to add a dilaton field to our previous action (\ref{eq:EMHD})
\be\label{eq:EMHD}
{\mathcal L} = \frac{1}{2 \k^2} \left(R + \frac{6}{L^2} - 2 \left(\nabla \Phi \right)^2 \right) - \frac{1}{4 e^2} e^{2 \a \Phi} F_{\mu\nu} F^{\mu\nu} - \frac{1}{\k^2} \left( |\nabla \phi - i A \phi|^2 + m^2 |\phi|^2 + V(|\phi|) \right) \,.
\ee
For ease of comparison we will take the same potential and mass for the charged scalar as we did in (\ref{eq:mV}) above. For simplicity we have chosen not to include a potential for the dilaton. Such a potential leads to interesting IR geometries \cite{Charmousis:2010zz}, whose interplay with superconducting instabilities deserves study at some future point. Taking the same ansatz for the fields as in (\ref{eq:spacetime}), with in addition the dilaton $\Phi(r)$,
there are now five equations of motion given by
\bea\label{eq:EOM2}
 r \phi'^2 + r \Phi'^2 + \frac{f'}{2f} + \frac{g'}{2g} + \frac{2}{r} + \g^2 r \, \frac{g h^2 \phi^2}{f} & = & 0 \,, \\
 r  \frac{h'^2 e^{2 \a \Phi}}{f} - \frac{f'}{f} + \frac{g'}{g} + \frac{6}{r} - r g (6 + 4 \phi^2 - u \phi^4) & = & 0 \,, \\
 h''  - h' \left(\frac{2}{r} + \frac{f'}{2f} + \frac{g'}{2g} - 2 \a \Phi' \right) - 2 \g^2 g h \phi^2 e^{- 2 \a \Phi} & = & 0 \,, \\
 \phi'' - \left(\frac{2}{r} - \frac{f'}{2f} + \frac{g'}{2g} \right) \phi' + \left(2g + \g^2 \frac{g h^2}{f} \right) \phi
 - u g \phi^3  & = & 0 \,, \\
\Phi'' - \left(\frac{2}{r} - \frac{f'}{2f} + \frac{g'}{2g} \right) \Phi' + \frac{\a h'^2 e^{2 \a \Phi}}{2 f} & = & 0 \,.
\eea

In the action (\ref{eq:EMHD}) we have chosen the dilaton to be massless, and it therefore corresponds to a marginal operator in the dual field theory with dimension $\Delta = 3$. We do not wish to turn on a coupling for this operator, and will therefore look for solutions to the equations of motion such that $\Phi \sim r^{3}$ near the asymptotically AdS boundary at $r \to 0$. With this extra tuning to perform, the numerics otherwise proceed as before.

In these theories the normal state, with $\phi = 0$, also has to be found numerically. However, the low temperature, near horizon normal state geometries may be found analytically \cite{Taylor:2008tg, Goldstein:2009cv}. In particular, the near horizon zero temperature metric has a Lifshitz form, and consequently one finds a specific heat scaling of
\be\label{eq:cn}
c_N \sim T^{2/z_N} \,, \qquad z_N = 1 + \frac{4}{\a^2} \,.
\ee
More specifically, the near horizon, $r \to \infty$, normal state solution at zero temperature takes the form
\be
\frac{1}{L^2}  ds^2 = - \frac{dt^2}{r^{2z_N}}  + g_L \frac{dr^2}{r^2} + \frac{dx^2 + dy^2}{r^2} \,, \quad A = \g \, h_L \frac{dt}{r^{2+z_N}} \,, \quad \Phi = \frac{2}{\a} \log r \,. \label{eq:sn}
\ee 
Where now $g_L = (1+z_N)(2+z_N)/6$ and $h_L^2 = (z_N-1)/(z_N+2)$.

Finite $z_N$ means that these theories have many fewer degrees of freedom in the normal state at low temperature compared to the models of the previous section. According to the entropy balance we might, therefore, anticipate that $T_c$ will be lower. This expectation is too na\"ive in general, as the entropy balance also depends on the low temperature degrees of freedom in the superconducting state, which we have not computed yet. However, it turns out to be true here.
Figure \ref{fig:tc} below shows $T_c/\mu$ as a function of $\g$ for $z_N = 1.5, 2$ and $3$, as well as for the locally critical model of the previous section. As previously, the critical temperature is independent of the quartic coupling $u$, as it only depends on the linearized equations of motion for the charged scalar. The plot is generated by looking for the temperature at which normalizable solutions to the linearized equation for the charged scalar field about the dilaton charged black hole background first exist.
\begin{figure}[h]
\begin{center}
\includegraphics[height=170pt]{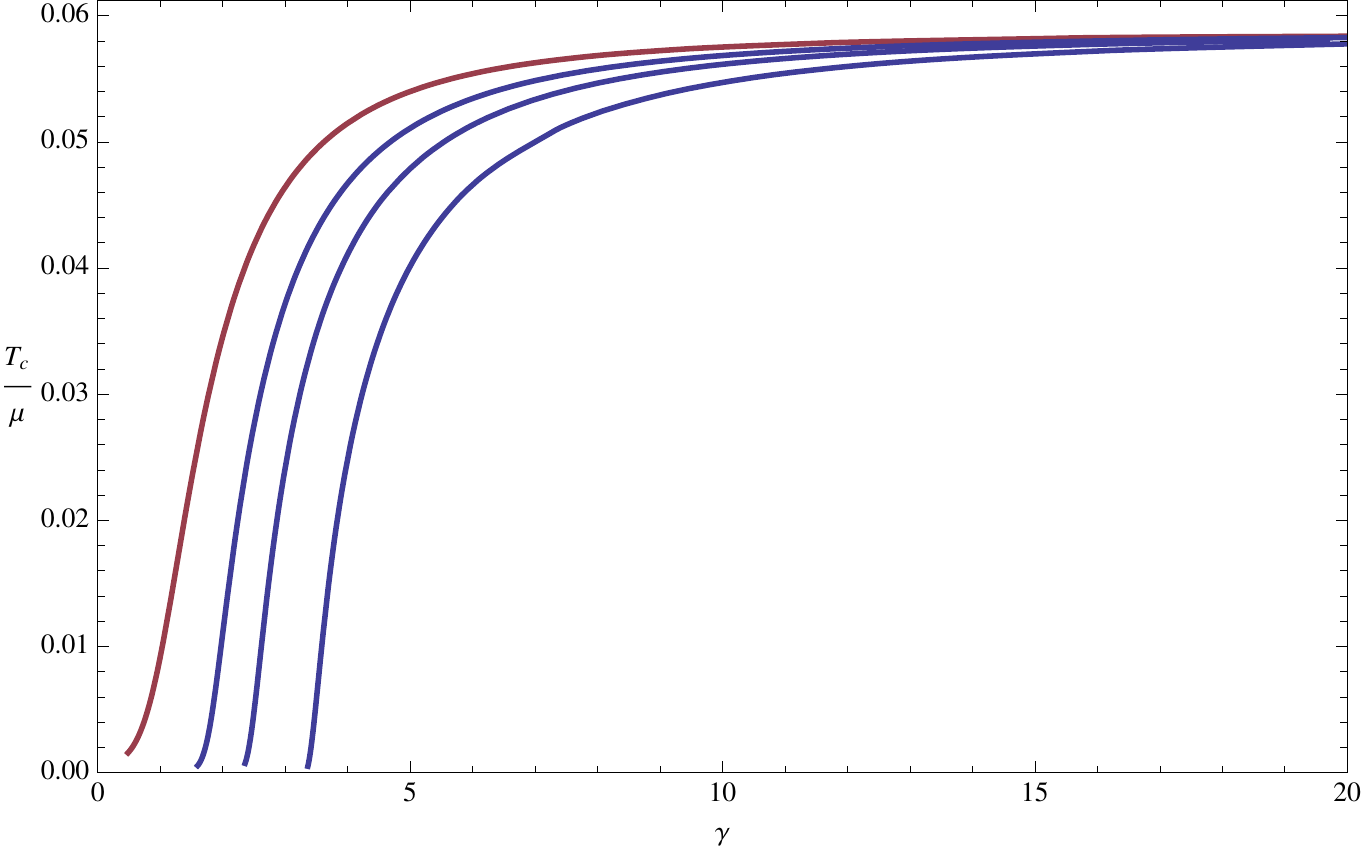}\caption{$T_c/\mu$ as a function of $\gamma$ for, from left to right, $z_N = \infty, 3, 2$ and $1.5$. \label{fig:tc}}
\end{center}
\end{figure}
In the plot we see firstly that at fixed $\g$, larger $z_N$ gives a larger $T_c$. We will see shortly that this is loosely analogous to larger gap giving larger $T_c$ in the BCS case. Secondly, we see that as $\g$ becomes large, $T_c/\mu$ tends to a constant value that is independent of $z_N$. This second observation can be understood mathematically from the fact that the large $\g$ limit is the probe limit, in which the backreaction of the matter fields on the metric becomes negligible \cite{Hartnoll:2008vx, Gubser:2008zu}. In the probe limit, $T_c$ is determined uniquely by the Abelian-Higgs theory in a neutral Schwarzchild-AdS background to be $T_c/\mu \approx 0.058$ \cite{Hartnoll:2008vx}, in agreement with figure \ref{fig:tc}.

It is interesting that, in this class of models, $T_c/\mu$ is bounded above by a rather small value. In the locally critical model, the critical temperature was obtained numerically as a function of $\g$ and the mass of the charged scalar field in \cite{Denef:2009tp}, but the quantity plotted there was $\g T_c/\mu$, so the universal limiting behavior (at fixed mass) was less immediate. While there is some ambiguity in the normalization of the charge density, it is natural to use (as emphasized in \cite{Denef:2009tp}) units in which charges are integrally quantized. This is the normalization we are using here.

In locally critical models, the existence of a superconducting instability at sufficiently low temperatures can be diagnosed from the behavior of the charged scalar field in the near horizon extremal $AdS_2 \times \R^2$ spacetime \cite{Hartnoll:2008kx, Denef:2009tp, Iqbal:2011in}. The charge of the scalar shifts its effective mass squared by the negative amount $g^{tt} A_t A_t$ \cite{Gubser:2008px}, which can help push the effective mass squared below the $AdS_2$ Breitenlohner-Freedman bound, leading to a superconducting instability. In fact, the mass we have taken, $m^2 L^2 = - 2$, already gives an instability of $AdS_2$ at $\g = 0$ where the scalar is neutral \cite{Hartnoll:2008kx}. We can see this in the $z_N = \infty$ curve in figure \ref{fig:tc}. In the Einstein-Maxwell-dilaton normal states being considered here, the IR effect of the charge of the scalar field is weaker. It is clear from the extremal near horizon solution (\ref{eq:sn}) that in these models, at zero temperature, $g^{tt} A_t A_t \to 0$ in the near horizon limit $r \to \infty$, and therefore the fact that the field is charged does not influence its IR mass squared. For the choice of mass $m^2 L^2 = - 2$ we can look for an IR instability by solving the
linearized equation of motion for the scalar in the near horizon Lifshitz background (\ref{eq:sn}). The result is (independently of $\g$)
\be
\phi \sim r^{1 + z_N/2 \pm \sqrt{4-z_N^2}/(2\sqrt{3})} \,.
\ee
A complex exponent is associated to an instability \cite{Denef:2009tp,Edalati:2011yv,Hartnoll:2011pp, Hartnoll:2011fn}. Therefore we see that there is an IR instability of the normal state against condensation of the scalar $\phi$ whenever $z_N > 2$. This is again consistent with the na\"ive notion that larger $z_N$ favors instability.
For $z_N < 2$ the instabilities shown in figure (\ref{eq:sn}) are not directly associated to the IR quantum critical regime, but depend on the full geometry. It is perhaps worthy of note that $z_N = 2$ has the same number of low energy degrees of freedom as a Fermi liquid, as measured by the specific heat, while $z_N < 2$ has fewer.

Before moving on to numerical results for these models, we can determine analytically the low temperature behavior of the superconducting phase. As in the previous model, the low temperature behavior will be controlled by the near horizon geometry of the extremal, zero temperature, solution. Previously we found a Lifshitz geometry (\ref{eq:lif}). The case with the dilaton is significantly more complicated because the dilaton and the charged scalar couple very differently to the Maxwell field. In particular, a purely Lifshitz type behavior combining the logarithmic running of the dilaton in (\ref{eq:sn}) and the constant value of the charged scalar in (\ref{eq:lif}) is not possible because the behavior of the Maxwell field is incompatible. Instead we take our cue from an insight in \cite{Hartnoll:2011pp}, where a similar technical challenge was faced for charged fermions. It is possible to find solutions where in the far IR the charged sector, i.e. the Maxwell field and charged scalar, are treated as a perturbation of a Lorentz-invariant IR spacetime. Thus we will find an emergent Lorentz invariant scaling with $z=1$ in the superconducting state at low temperatures.

One can solve the equations of motion (\ref{eq:EOM2}) with the following expansion as $r \to \infty$:
\bea
f & = & \frac{1}{r^2} \left(1 + f_1 r^{3-a} + \cdots \right) \,, \label{eq:aa3} \\
g & = & \frac{3 u}{2 + 3 u} \frac{1}{r^2} \left(1 + g_1 r^{3-a} + \cdots \right) \,, \\
h & = & h_1 r^{(1-a)/2} + \cdots \,, \\
\phi & = & \sqrt{\frac{2}{u}} +p_1 r^{3-a} + \cdots \,, \\
\Phi & = & - \frac{1}{2 \a} \log \frac{(a^2-1)(2+3 u)}{48 \g^2} + P_1 r^{3-a} + \cdots \,. \label{eq:bb3}
\eea
The exponent $a$ is not fixed; in order for this expansion to be valid we must have $a > 3$. The constants $\{f_1,g_1,h_1,p_1,P_1\}$ are easily determined but we will not need them. The above expansions describe an emergent IR $AdS_4$ spacetime in the superconducting state at zero temperature. This a familiar phenomenon, e.g. \cite{Gubser:2008wz, Gubser:2009gp, Gauntlett:2009dn}. Our experience here suggests that the emergence of an $AdS_4$ may be generic in models with both dilatons and charged scalar fields. The incompatible couplings of the two scalars to the Maxwell field effectively force the Maxwell field to go to zero in the IR. This indicates that the charge density operator is becoming irrelevant, and hence Lorentz symmetry re-emerges.

The emergent Lorentz symmetry implies that the specific heat at low temperature in the superconducting state will go like
\be
c_S \sim T^2 \,.
\ee
These models therefore have in common with the previous locally critical normal state models that the critical modes are not entirely gapped in the superconducting state. An interesting difference is the following:
In the model with a dilaton, the normal state has a tunable $z_N$ while the superconducting state has $z=1$ fixed. The locally critical model had $z_N = \infty$ fixed but $z$ tunable. More interesting, however, is that in both cases we have
\be\label{eq:zzn}
z < z_N \,.
\ee
While, like most of our observations, this inequality does not follow as a necessity from entropy balance, it is nicely in agreement with a simple-minded application of the need to balance the degrees of freedom; the superconducting phase has more degrees of freedom than the normal phase at higher temperatures, just below $T_c$, and so should have fewer degrees of freedom at lower temperatures. A smaller $z$ indeed implies fewer low energy degrees of freedom according to (\ref{eq:clowT}). It would be interesting to see how robustly the inequality (\ref{eq:zzn}) holds in holographic models in which both the normal and superconducting dynamical critical exponents can be tuned. Models in which the normal state Lifshitz geometry is achieved via charged fermions rather than dilatonic scalars  \cite{Hartnoll:2009ns, Hartnoll:2010gu} may provide a framework to address this question.

\begin{figure}
\begin{center}
\includegraphics[height=540pt]{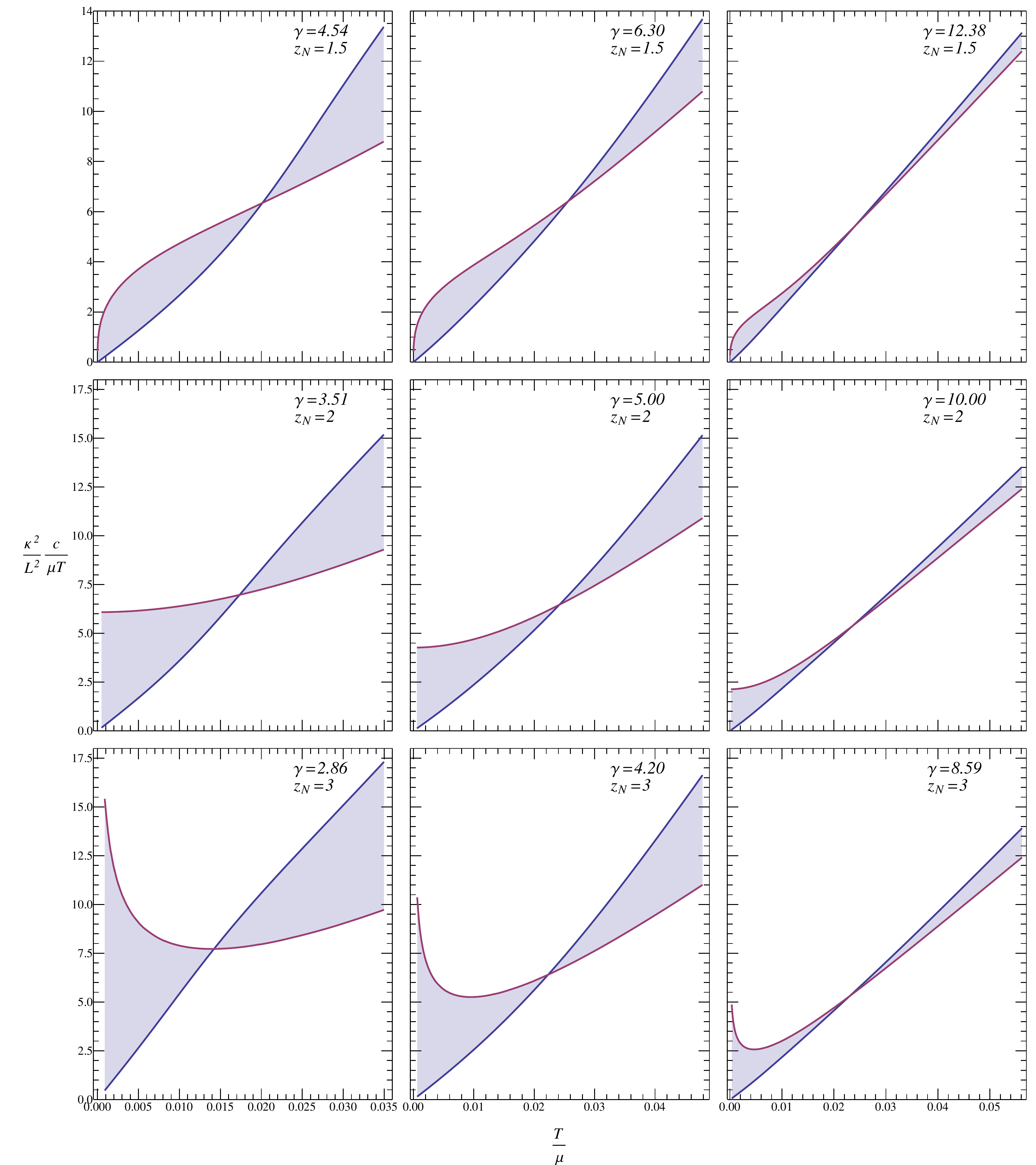}\caption{Specific heat over temperature as a function of temperature
for the normal (red) and superconducting (blue) phases below $T_c$. These plots are for models without a ground state entropy density. Three values of $T_c/\mu$ and $z_N$ are shown. The parameter $\g$ has been chosen in each case to make the three values of $T_c/\mu$ equal. The shaded region between the blue and red curves integrates to zero area.
 \label{fig:covert2}}
\end{center}
\end{figure}

In figure \ref{fig:covert2} we plot the specific heat over temperature for the normal and superconducting states
for temperatures below the critical temperature. We chose the three values of $z_N = 1.5, 2, 3$ in order to illustrate cases in which the specific heat over temperature goes to zero, a constant and infinity at low temperatures, according to (\ref{eq:cn}), respectively. These behaviors are clearly visible in the plots. We have already noted that at fixed $\g$, the critical temperature increased with $z_N$. In figure \ref{fig:covert2} we have chosen $\g$ so that the three values of $T_c/\mu$ shown are the same for each value of $z_N$ and furthermore comparable to the critical temperatures we plotted for the locally critical case.

The anticipated linear behavior of $c_S/T$ at low temperatures is also clearly visible in the plots. We have checked that the numerics reproduce in detail the low temperature behavior of equations (\ref{eq:aa3}) -- (\ref{eq:bb3}).
In fact, the curves for $c_S/T$ are essentially featureless and decrease monotonically from their values at $T_c$. These plots are closer to the BCS curve of figure \ref{fig:bcsovert} than the locally critical models. The main difference with the BCS case, beyond the fact that the ordered state is not gapped here, is that the shape of the curves changes as $T_c$ is increased; the jump in specific heat at the phase transition decreases with increasing $T_c$. More generally, the two curves come closer as $T_c$ is increased, consistent with the fact, mentioned above, that in the large $\g$ limit the charged sector of theory becomes an increasingly negligible `probe' sector that does not backreact on the neutral degrees of freedom. One way to phrase this is that $\g$ controls the static charge susceptibility of the theory, which is becoming small.

\section{Discussion}

In this paper we have used the notion of entropy balance to relate qualitative features of the normal and superconducting phases in various models of holographic superconductivity. As part of this investigation we have introduced new holographic superconductors in which the critical normal state from which the superconductivity emerges has finite dynamical critical exponent $z_N$.
While entropy balance itself is a rigorous statement, we have emphasized that the nature of the relations we ascribe to it is not one of logically necessary connections, but rather of tendential relations that one might `na\"ively' but reasonably expect to hold more often than not. The main observations connected to entropy balance that we have made for the models considered in this paper are that (a) $z < z_N$, (b) larger $z_N$ leads to a larger $T_c$ if other parameters are kept fixed, (c) $c_S/T$ starts decreasing below $T_c$ in models without a ground state entropy, but increases in models with a ground state entropy density, (d) in models with a ground state entropy density and a low critical temperature, critical scaling in the superconducting state is pushed down to very low temperatures. The simplest way in which these statements can fail to hold is if the jump in specific heat at the critical temperature varies in such a way that it cancels out the effects of low temperature criticality in the entropy balance formula. It is of interest perhaps to understand more deeply the physics controlling this jump.

Our observations cannot be applied directly to either the nematic or ferromagnetic systems that we cited in the introduction as motivation for this work. The ferromagnetic systems have first order phase transitions. The nematic phase in Sr$_3$Ru$_2$O$_7$ does have a second order transition, but the ordered phase develops out of an anomalous Fermi liquid normal state in which $c_N/T$ is logarithmically increasing as the temperature is lowered. If this logarithm is extrapolated down to zero temperature, then it is able to precisely entropically balance the anomalous increase in the specific heat of the ordered phase, without recourse to a ground state entropy density in the normal phase \cite{andy}. This is ultimately tied to the fact that nematicity following from a Pomeranchuk instability, unlike superconductivity, leaves the existence of a Fermi surface intact and therefore allows many more low temperature degrees of freedom in the ordered phase. An increasing specific heat below $T_c$ in a superconductor would be more dramatic.

Heavy fermion superconductors are another class of systems with unconventional normal and superconducting states, in which entropy balance has been explored experimentally \cite{heavy0,heavy1}. The heavy fermion materials exhibit power law specific heat at low temperatures in the superconducting state, and thus, like the models we have considered, have more degrees of freedom at low temperature than gapped BCS superconductors. Unlike the holographic models considered in this paper, which have scalar condensates, the extra degrees of freedom in the heavy fermions are naturally associated with nodal excitations of the non-s-wave Cooper pairing.  Somewhat similarly to Sr$_3$Ru$_2$O$_7$, these extra degrees of freedom are entropically balanced out with additional degrees of freedom in the non-Fermi liquid normal state \cite{heavy0,heavy1}.\footnote{The way in which entropy balance works out in these materials suggests that $T_c$ would be increased if the non-s-wave condensates were to emerge from a strict Fermi liquid normal state, without a logarithmic enhancement of the low temperature degrees of freedom.} In CeCoIn$_5$ there appears, in addition to the power law in temperature, a rather spectacular surplus of very low temperature degrees of freedom in the superconducting phase \cite{heavy2}. The entropic connection between unconventional normal states and unconventional superconductivity in these materials is precisely the type of physics we have attempted to get a computational handle on in this paper. It may be possible to make a tighter connection to the materials if the holographic emergence of superconductivity from a fractionalized phase can be related to, for instance, spin density wave physics, in the spirit of e.g. \cite{doublon}.

Specific heat measurements also exist for the cuprates and show some similar features, e.g. \cite{hightc}. In this case, extracting sharp information from the measurements is complicated by the need to subtract a phonon background, due to the higher temperatures involved.

Beyond the models we have considered here, there are other holographic contexts where the interplay of superconductivity and quantum criticality may be usefully studied. Quantum critical normal states with $z_N$ finite can be constructed using `electron stars' \cite{Hartnoll:2009ns, Hartnoll:2010gu} rather than dilatonic fields. These are qualitatively different in that the asymptotic electric flux is sourced by charged fermions in the bulk rather than emanating from behind a horizon \cite{Hartnoll:2011fn}. Thus any superconducting instability in this case is not coincident with exit from a fractionalized phase, but corresponds to the re-arrangement of confined charge-carrying `quark' degrees of freedom  from fermionic to bosonic `mesons' \cite{Hartnoll:2011fn}. This phenomenon has been studied for neutral scalar instabilities in \cite{Edalati:2011yv}, but not yet for charged scalars.

A further way to probe the interrelation between deconfined critical degrees of freedom and superconductivity is to incorporate the physics of confinement. Holographic models have been developed that exhibit confining as well as superconducting phases \cite{Nishioka:2009zj, Horowitz:2010jq}. It is of interest to extend our computations to include these effects. In particular, following a possibly overly na\"ive application of entropy balance, one would like to confirm the extent to which gapping out the critical modes in the normal state causes a reduction in the critical temperature.

\section*{Acknowledgements}

It is a pleasure to thank Steve Kivelson, Andy Mackenzie, Subir Sachdev and Jan Zaanen for bringing the notion of entropy balance to our attention and for helpful discussions on the topic. We have also benefitted greatly from discussions with Andrew Green and Robert B. Mann. The work of S.A.H. is partially supported
by a Sloan research fellowship. R.P. is supported by the Natural Sciences and Engineering Research Council
of Canada.

\end{document}